\documentclass[11pt,twoside]{article}

\usepackage{asp2014}

\aspSuppressVolSlug
\resetcounters

\begin{document}

\title{Probabilistic photo-z machine learning models for X-ray sky surveys}

\author{Viktor~Borisov,$^{1,2}$ Alex~Meshcheryakov,$^1$ Sergey~Gerasimov,$^2$ and RU eROSITA catalog group$^1$}
\affil{$^1$Space Research Institute of the Russian Academy of Sciences, Moscow, Russia \email{victor.d.borisov@gmail.com}}
\affil{$^2$Lomonosov Moscow State University Faculty of Computational Mathematics and Cybernetics, Moscow, Russia}

\paperauthor{Viktor~Borisov}{victor.d.borisov@gmail.com}{https://orcid.org/0000-0002-5974-5539}{Lomonosov Moscow State University Faculty of Computational Mathematics and Cybernetics; Space Research Institute of the Russian Academy of Sciences}{}{Moscow}{Moscow}{119991}{Russia}
\paperauthor{Alex~Meshcheryakov}{mesch@cosmos.ru}{https://orcid.org/0000-0002-7543-4183}{Space Research Institute of the Russian Academy of Sciences}{High Energy Astrophysics}{Moscow}{Moscow}{117997}{Russia}
\paperauthor{Sergey~Gerasimov}{sergun@gmail.com}{}{Lomonosov Moscow State University}{Faculty of Computational Mathematics and Cybernetics}{Moscow}{Moscow}{119991}{Russia}
\paperauthor{RU eROSITA catalog group}{}{}{Space Research Institute of the Russian Academy of Sciences}{High Energy Astrophysics}{Moscow}{Moscow}{117997}{Russia}



  
\begin{abstract}

Accurate photo-z measurements are important to construct a large-scale structure map of X-ray Universe in the ongoing SRG/eROSITA All-Sky Survey.  We present machine learning Random Forest-based models for probabilistic photo-z predictions based on information from 4 large photometric surveys (SDSS, Pan-STARRS, DESI Legacy Imaging Survey, and WISE). Our models are trained on the large sample of $\approx$580000 quasars and galaxies selected from the SDSS DR14 spectral catalog and take into account Galactic extinction and uncertainties in photometric measurements for target objects. On the Stripe82X test sample we obtained photo-z accuracy for X-ray sources: $NMAD=0.034$  (normalized median absolute deviation) and $n_{>0.15}=0.088$ (catastrophic outliers fraction), which is almost $\sim2$ times better than best photo-z results available in the literature.

\end{abstract}

\section{Introduction}
On July 13, 2019 the SRG X-ray observatory
was launched from the Baikonur cosmodrome. On Dec. 8th, 2019 SRG started its first All-Sky Survey, which will consist of 8 repeated six month long scans of the entire sky. eROSITA telescope \citep{2020arXiv201003477P} onboard SRG operates in the soft X-ray band (0.3–8\,keV) and will detect $\sim3$ millions X-ray AGNs at the end of survey. In order to construct a large-scale structure map of X-ray Universe with eROSITA, accurate measurements of cosmological redshifts for extragalactic X-ray sources (mostly quasars) are needed.

Redshift measurement methods \citep{2019NatAs...3..212S} can be divided into spectroscopic (spec-z, $z_{sp}$) and photometric (photo-z, $\hat{z}_{ph}$). Spec-z's are time consuming task for faint optical objects ($r\gtrsim22^{mag}$). On the other hand, photo-z measurements can be based on data from modern large photometric sky surveys, it is much cheaper in observational resources than spec-z but also less accurate. 

In this work we present machine learning models for X-ray sources probabilistic photo-z predictions, based on photometric data from 4 modern sky surveys (SDSS, Pan-STARRS1, DESI Legacy Imaging Survey, and WISE).

\section{Data}\label{sec:data}

We use photometric data from SDSS DR14 \citep{2018ApJS..235...42A}, Pan-STARRS1 DR2 \citep{2018AAS...23110201C}, DESI LIS DR8 \citep{2019AJ....157..168D} and WISE \citep{2010AJ....140.1868W} sky surveys (WISE forced photometry is taken from DESI LIS).

Firstly, for all used photometric surveys we calculated hyperbolic magnitudes ($mag$) from object fluxes ($flux$) and uncertainties ($\sigma_{flux}$): 
\begin{equation}\label{eq:asinh}
    mag = \Bigg[asinh\Bigg(\frac{flux}{2 \times \sigma_{flux}}\Bigg) + \log(\sigma_{flux})\Bigg] \times \Bigg(\frac{-2.5}{\log 10}\Bigg) ~.
\end{equation}
We derived the following set of photometric features:
\begin{itemize}
    \item 5 PSF and 5 model magnitudes ($u$, $g$, $r$, $i$, $z$) from SDSS and related 15 colors (i.e. $g_{psf}-g_{model}$, $g_{psf}-r_{psf}$, etc; colors like $g_{model}-r_{model}$ were not used),
    \item 5 PSF and 5 Kron magnitudes ($g$, $r$, $i$, $z$, $y$) from Pan-STARRS and related 15 colors (like with SDSS),
    \item 3 model magnitudes ($g$, $r$, $z$) from DESI LIS and related 6 colors,
    \item 2 WISE magnitudes ($w1$, $w2$) and 1 related color ($w1-w2$), 
    \item 10 colors between WISE and SDSS magnitudes, 10 colors between WISE and Pan-STARRS Kron magnitudes, and 6 colors between WISE and DESI LIS $g$, $r$, $z$ magnitudes,
    \item 3 colors between DESI LIS and SDSS model mags in  $g$, $r$, $z$ bands.
\end{itemize}
Magnitudes and colors were corrected for Galactic extinction by using $E(B-V)$ estimates from DESI LIS and $A/E(B-V)$ coefficients from \citep{2011ApJ...737..103S} ($R_V=3.1$ was adopted).

Our \textbf{training dataset} contains 449751 quasars from SDSS DR14q \citep{2018A&A...613A..51P} and 136428 galaxies from SDSS DR14 (a subsample of galaxy catalog filtered to approximate the distribution of spectroscopic subclasses for optical counterparts of X-ray sources). To increase the number of distant ($z > 5$) quasars in the dataset, we added all sources from \textbf{VHzQs sample} \citep{2020MNRAS.494..789R}.

\section{Method}\label{sec:method}

Our aim was to estimate conditional redshift distribution $p(z|x)$ for each target object with photometric features $x$.
We use Random Forest (RF) model, \citep{2001MachL..45....5B,JMLR:v7:meinshausen06a}, which is considered by many authors among the most accurate ML algorithms for photo-z measurements of galaxies \citep{2020MNRAS.499.1587S,2020arXiv200912112E} and X-ray quasars \citep{2018AstL...44..735M}. We used RF ensemble predictions in combination with gaussian Kernel Density Estimation (gKDE), to obtain $p(z|x)$. RF+gKDE model allows one to calculate photo-z point estimate $\hat{z}_{ph} = \arg\max_z p(z|x)$, confidence intervals, and $zConf = \int_{\delta z_{norm} < 0.06} p(z|x)~dz$.

At the prediction stage we take into account uncertainties in photometric fluxes of the target object, by perturbing  fluxes (according to given uncertainties) for each regression tree in the forest.

\section{Results}\label{sec:eval}

We trained various photo-z models that use features from different combinations of photometric surveys: SDSS + WISE, Pan-STARRS + WISE, Pan-STARRS + DESI~LIS + WISE and SDSS + Pan-STARRS + DESI~LIS + WISE. 

We use the \textbf{Stripe82X} sample of X-ray sources with known spectroscopic redshift \citep{2017ApJ...850...66A} in order to evaluate accuracy of photo-z models point predictions. Standard metrics for point predictions were used: 
$NMAD=1.4826\times median(|\delta z_{norm}|)$) --- normalized median absolute deviation,  
$n_{>0.15}$ --- fraction of catastrophic outliers with $\delta z_{norm}>0.15$,
where $\delta z_{norm} = \frac{\hat{z}_{ph} - z_{sp}}{1+z_{sp}}$.

Comparison of photo-z models on the Stripe82X sample are shown in Table \ref{tab:eval}. As one can see, our photo-z models demonstrate better accuracy than State-Of-The-Art methods (template photo-z models \citep{2017ApJ...850...66A} and neural network photo-z model \citep{2019MNRAS.489..663B}) on the same test sample. Our most accurate photo-z model (based on photometric data from SDSS, PanSTARRS, DESI LIS, WISE surveys) outperforms \citep{2017ApJ...850...66A} and \citep{2019MNRAS.489..663B} results by a factor of $\approx2$.

\begin{table}[!ht]
    \caption{Comparison of photo-z models on Stripe 82X test sample of X-ray sources with known spectroscopic redshift.}\label{tab:eval}
    \smallskip
    \begin{center}
    {\small
    \begin{tabular}{llc}
        \tableline
        \noalign{\smallskip}
        Model & $NMAD$ & $n_{>0.15}$\\
        \noalign{\smallskip}
        \tableline
        \noalign{\smallskip}
        SDSS + WISE & 0.047 & 0.119 \\
        Pan-STARRS + WISE & 0.056 & 0.152 \\
        Pan-STARRS + DESI LIS + WISE & 0.045 & 0.115 \\
        DESI LIS + WISE & 0.072 & 0.176 \\
        SDSS + Pan-STARRS + DESI LIS + WISE & \textbf{0.034} & \textbf{0.088} \\
        \noalign{\smallskip}
        \tableline
        \noalign{\smallskip}
        Template model \citep{2017ApJ...850...66A} & 0.065 & 0.170 \\
        Neural network \citep{2019MNRAS.489..663B} & 0.066 & 0.156 \\
        \noalign{\smallskip}
        \tableline
    \end{tabular}
    }
    \end{center}
\end{table}

\section{Conclusion}\label{sec:conclusion}
We study the problem of measuring photometric redshifts
(photo-z) of extragalactic X-ray sources using machine learning techniques. The proposed photo-z models based on Random Forests show accuracy (up to 2 times) better than current SOTA results in the literature (on Stripe82X field). The main accuracy improvement comes from using a large training sample ($\sim$600k objects) with features from modern wide photometric surveys. First optical spectroscopic observations of eRosita sources show that the proposed photo-z models are effective in the ongoing search for distant X-ray quasars (see e.g. \citep{2020MNRAS.497.1842M,2020AstL...46..149K,2020AstL...46..429D}).

The presented photo-z models are integrated into the SRGz system designed to construct a three-dimensional map of X-ray sources on the Eastern Galactic Hemisphere of the SRG/eRosita All-Sky survey. The SRGz system is developed in the science working group of RU eROSITA consortium on X-ray source detection, identification, and eROSITA source catalog in the High Energy Astrophysics Department at Space Research Institute of the Russian Academy of Sciences.

\bibliography{P5-215}


\end{document}